\documentclass[sigconf]{acmart}

\AtBeginDocument{%
  \providecommand\BibTeX{{%
    \normalfont B\kern-0.5em{\scshape i\kern-0.25em b}\kern-0.8em\TeX}}}

\copyrightyear{2021}
\acmYear{2021}
\setcopyright{acmcopyright}
\acmDOI{}

\acmConference[Pre-print]{1}{June}{2021}
\acmBooktitle{}
\acmPrice{}
\acmISBN{}

\setcopyright{none}
\renewcommand\footnotetextcopyrightpermission[1]{}
\begin{document}

\title{Identifying Roles, Requirements and Responsibilities\\in Trustworthy AI Systems}

\author{Iain Barclay}
\email{BarclayIS@cardiff.ac.uk}
\affiliation{%
  \institution{School of Computer Science \& Informatics,\\Cardiff University}
  \streetaddress{Queen's Buildings}
  \city{Cardiff}
  \country{UK}
  \postcode{43017-6221}
}
\author{Will Abramson}
\email{will.abramson@napier.ac.uk}
\affiliation{%
  \institution{Blockpass ID Lab, School of Computing, \\Edinburgh Napier University}
  \city{Edinburgh}
  \country{UK}
  \postcode{43017-6221}
}



\begin{abstract}
Artificial Intelligence (AI) systems are being deployed around the globe in critical fields such as healthcare and education. In some cases, expert practitioners in these domains are being tasked with introducing or using such systems, but have little or no insight into what data these complex systems are based on, or how they are put together. In this paper, we consider an AI system from the domain practitioner's perspective and identify key roles that are involved in system deployment. We consider the differing requirements and responsibilities of each role, and identify a tension between transparency and privacy that needs to be addressed so that domain practitioners are able to intelligently assess whether a particular AI system is appropriate for use in their domain.
\end{abstract}


\maketitle

\section{Introduction}
Artificial Intelligence (AI) systems can be considered as deployable instances of Machine Learning (ML) models, provided with user interfaces and other supporting infrastructure. These  systems are designed and implemented for a specific purpose within a domain by engineers and AI experts, to be used by practitioners working within the domain to augment their abilities. The practitioner's work can be of critical importance, with life-changing impact - across fields as diverse as education, healthcare, finance and food production, and AI is typically adopted to assist with complex decision making. AI systems have a heavy reliance on data, both for training and testing the ML models that provide the core functionality of the systems, and for making operational analysis in the domain. Practitioners will generally interact with an AI system via an app or web browser interface, so that they can provide the system with data and view any results from the system. Actual deployment and use of such AI systems in real-world environments is beginning to highlight ``last mile''\cite{cabitza2020bridging} AI deployment challenges.

The motivation for this paper comes from the draft publication of UNICEF's Policy Guidance on AI for Children\cite{unicef2020}, which states that: ``Data equity and representation of all relevant children for a particular AI system, including children from different regions (including rural communities), ages, socioeconomic conditions and ethnicity's, is essential to protect and benefit children. \emph{For example, in the case of data-driven health care, children's treatment or medication should not be based on adults' data since this could cause unknown risks to children's health.} [emphasis added]''. This paper seeks to address this requirement by considering the constructs that would need to be in place so that a party with responsibility in the deployment environment --- a domain practitioner (DP), a manager, inspector or other responsible role --- could seek and gain assurance that an AI system is appropriate for use within a given context. For example a system designed to diagnose medical conditions in children was indeed not based on adults' data. The notion of what is \emph{appropriate} will vary from case to case, and in some situations may involve regulatory authorities, best practice or simply good judgement from an experienced practitioner. In many scenarios a citizen, or their guardian, has no alternative than to put their faith in a DP to have performed some degree of due diligence and chosen their tools with care. This responsibility extends to the use of AI systems, which places an onus upon DPs to be able to provide citizens with assurance on the suitability of a system. To provide this, the practitioner themselves must be able to develop a level of confidence in the supplier of the system, and in the suitability of the components and processes used to build the system.

In order to bring insight into the challenges faced in determining whether an AI system under consideration is appropriate, this paper adopts the DP's perspective, and considers a hierarchy of different roles that they are reliant upon for the manifestation of an AI system. Having identified representative roles, each is considered in turn to identify their responsibilities and their requirements, or goals, that place demands on others in the ecosystem as AI systems go from data, through model development and systems integration to potential deployment in the field.

The contribution of this paper is the development of a framework for understanding AI systems from the point-of-view of a DP, which in turn highlights the different roles, requirements and responsibilities of  participants in the ecosystems that lead to the ultimate deployment and on-going maintenance of such systems. The framework provides a means to identify potential tensions between stakeholders in the system, which has implications for the design of AI systems in practice.
The remainder of this paper is structured as follows: Section  \ref{chapter:related} introduces related work, Section \ref{chapter:overview} provides an analysis of different roles and responsibilities in an AI system, which is visualized in Section \ref{chapter:modelling}. A discussion on the implications for AI system design is provided in Section \ref{chapter:discussion}, prior to Conclusions which are offered in Section \ref{chapter:conclusions}.

\begin{figure*}[t]
\begin{center}
\includegraphics[width=0.8\textwidth]{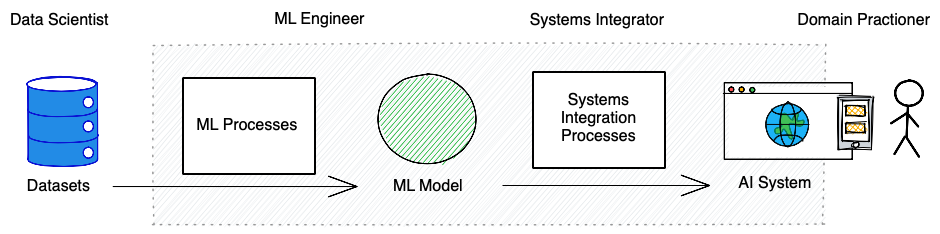}
\end{center}
\vspace{-10pt}
\caption{Components of an AI system, adapted from Renggli, et al.~\cite{renggli2021data}}
\label{fig:aipipeline}
\vspace{-5pt}
\end{figure*}

\section{Related Work}
\label{chapter:related}

The research field of Explainable AI\cite{bhatt2020explainable} seeks to provide insight into the decisions made by AI and ML systems to a range of different stakeholders. Researchers have identified different roles that might require explanations of how an AI system has made a decision. Tomsett et al.~\cite{tomsett2018interpretable}, for example, identified six classes of explanation recipients: system creators, system operators, executors making a decision on the basis of system outputs, decision subjects affected by the executor's decision, the data subjects whose personal data is used to train a system, and system examiners (e.g., auditors or ombudsmen). Building on this work, Preece et al.~\cite{preece2018stakeholders}, examined stakeholder needs for explanations and considered Users as one of four interested groups, alongside Theorists, Ethicists and Developers, and conclude that ``the most influential of our four stakeholder communities is the users''.




Explainable AI covers a wide scope, and providing information and ``audit trails''\cite{brundage2020toward} on the parts and processes that have led to the generation of a model is recognised as a valuable component. Nonetheless, the decision of a DP to adopt an AI system and apply it to their domain, is ultimately a decision they make to place trust in the system within the context that they are applying it. The DP is unable to know with certainty how the system was constructed, what data was used and which actors were involved in it's construction. As such the choice to adopt and use an AI system is risky, and the DP is responsible for the actions they take as a result of the information they receive from the system. As risk increases, the act of placing trust becomes harder to justify \cite{nickel2012risk}. Those who place trust intelligently, seek out evidence to provide assurance that their judgement is appropriate. This evidence must be judged for it's accuracy and relevance, which requires the ability to attribute claims to their sources whose authority and trustworthiness must also be assessed \cite{o2002question,carter2020ethics}. 

Assessing trustworthiness in complex multi-actor sociotechnological systems presents many challenges. Hawlitscheck, et al.\cite{hawlitschek2018limits} surveying trust within the sharing economy found the literature on trust predominantly focuses on human interactions as opposed to the computational systems that facilitate these interactions. This work introduce the term \emph{Trust Frontiers}, to describe these boundaries between human interaction and the underlying technical systems that are increasingly a part of these interactions. This terminology is adopted in our analysis, as we consider the various interfaces where trustworthiness needs to conveyed between human and algorithmic components in AI systems.


\section{AI System Overview}
\label{chapter:overview}
For the purposes of the discussion, an AI system is considered to be a deployable instance of an ML model, which has been packaged with a suitable user interface, to meet the needs of a particular customer or end user. The production of an AI system includes processes which lead to the development of an ML model from one or more datasets, increasingly this chain of processes is referred to as the ``MLOps Pipeline''\cite{renggli2021data}. ML models are then integrated with supporting interfaces and application logic through software development processes with the aim of creating a usable, domain-specific AI system solution for deployment in the field. This production pipeline yields the components, and machine learning and software integration processes illustrated in Figure~\ref{fig:aipipeline}. This is a somewhat traditional view, in that it begins with the input --- the datasets --- on the left-hand side, and progresses through time to the user's domain on the right-hand side.

\subsection{AI System Hierarchy}

\begin{figure*}[ht]
\begin{center}
\includegraphics[width=0.5\textwidth]{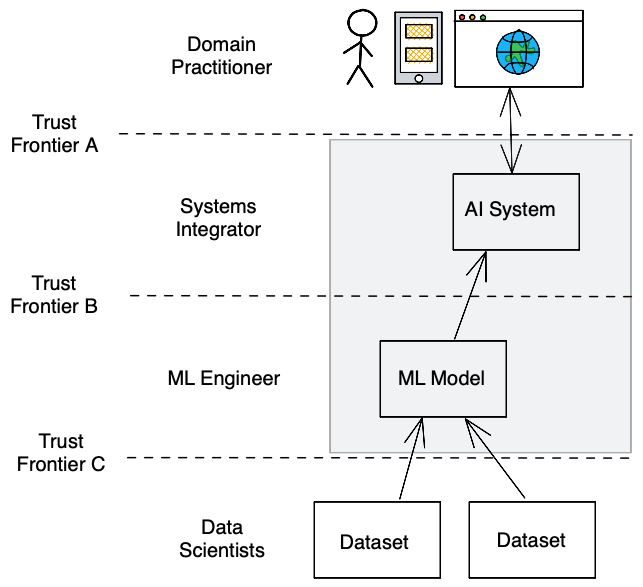}
\end{center}
\vspace{-10pt}
\caption{Roles and Trust Frontiers in AI Systems}
\label{fig:roles}
\vspace{-5pt}
\end{figure*}

Figure~\ref{fig:roles} re-orientates the production pipeline into a hierarchical view, to highlight the dependencies between the parties involved and the components they provide. The illustrated components, such as datasets and ML models, can be used across multiple AI systems in different combinations, and deployed in different domains. This hierarchy will be reproduced across every AI system, with some of the datasets or the ML models shown here appearing in other systems too. The boundaries between the roles shown in Figure~\ref{fig:roles} are identified as Trust Frontiers\cite{hawlitschek2018limits}. These boundaries denote the interfaces where information flow is required between roles, where actors rely upon the integrity of others in the hierarchy. Note that in some deployments of AI systems individual actors in the ecosystem will take on several of the identified roles, yet in other cases there may be no direct relationship between the parties or any knowledge of the other actors.

\subsection{Domain Practitioners}
The DP is the end user of an AI system, which is an implementation that uses the ML model to provide analysis on a particular problem in a specialised domain, for example, children's healthcare. The DP, and stakeholders related to the DP\footnote{For brevity, the nomenclature of Domain Practitioner is used throughout this paper, but this role is intended to include any party or stakeholder with responsibilities for the selection, procurement or application of the AI system in its domain environment.}, are unlikely to have AI/ML expertise, but have a duty of care to their patients, clients or customers - the people ultimately impacted by decisions informed by the AI system.

\subsubsection{Responsibilities}
The DP has a responsibility to ensure that the AI system is fit for purpose in their domain. This may mean that it meets requirements set down in any local laws, or rules set by regulators or professional bodies. Within the scope of this paper, it is considered that the DP may be required to seek assurance that the underlying data and processes used in the development of the AI system are \emph{appropriate} for use. 

\subsubsection{Appropriate ML Model and Data Requirement}
The DP is reliant on their Systems Integrator (SI) providing them with a robust, reliable system and that the SI acted with integrity and good judgement when selecting ML models for use in the AI system. There is a further dependency on the ML Engineer (MLE) providing sufficient, trustworthy information on the constituent components of the ML model to the SI, and, in turn on the MLE having received reliable information from data suppliers, and being able to provide satisfactory evidence of this. Referring to Figure~\ref{fig:roles}, the DP seeks information through Trust Frontier A, but is also reliant on other parties deeper in the hierarchy developing relationships and being able to provide evidence across their own Trust Frontiers. 

\subsubsection{Transparency Requirement}
In order to determine the suitability of the AI system for use in their environment the DP may have a need for some degree of visibility or transparency into the underlying components and processes which built it, so that they can form a judgement according to their principles. This need for transparency will potentially be in conflict with the other parties desires for privacy due to desires to protect commercial information or in some cases legal requirements placed upon other parties can also prevent them from providing information on individuals, such as employees involved in system development.

\subsection{The Systems Integrator}
The SI\footnote{Many aspects of this role include software engineering tasks, but the Systems Integrator moniker was chosen to highlight the need to integrate or combine many components to deliver the solution} is the role responsible for providing an AI system suitable for use by the DP.
The SI role can be considered to be ``responsible for designing and integrating externally supplied product and service components into a system for an individual customer''\cite{davies2007organizing}. The SI will typically provide an interface through which the DP can provide case data in the domain field and the results from the underlying ML model will be displayed. This interface will often be an app or a web interface, but could use other technologies such as voice or physical sensor devices. The SI will select one or more ML models, either from their own organisation or from a supplier. The SI will have many other tasks to perform to deliver a reliable AI system. They may work directly with the DP to meet agreed requirements or deliver their AI system as a service that can be found and used by any DP, without a direct relationship necessarily being in place between the SI and DP.

\subsubsection{Responsibilities}
The SI has a responsibility to ensure that the AI system they deliver is robust and reliable, and available when the DP needs it. There is a strong reliance on the SI choosing suitable ML models, and performing necessarily due diligence in this selection process. The SI also has a responsibility to use any chosen ML model correctly, and in accordance with any specifications laid down by the ML model providers.

\subsubsection{Appropriate ML Model and Data Requirement}
The degree to which an SI can determine the suitability of an ML model and the datasets used to develop the models they consider for adoption in AI systems depends on information that is available about the model and the data sets, and their perception of the trustworthiness of the parties providing documentation and evidence in support, through Trust Frontier B in Figure~\ref{fig:roles}. Such considerations can include a need to know the design parameters for the model, and the conditions under which the data was generated. The required levels of trustworthiness can arise from knowledge of the party who created and shared the model, and to some degree can be delivered through documentation that is supplied with the model. 

\subsubsection{Privacy Requirement}
Whilst the SI may be keen to engage with the DP to provide assurances on the qualities of their product, there is some information that they may be unwilling to share, or indeed unable to share. Such information might include sensitive commercial relationships or details of algorithms, where the SI may have commercial motivation to restrict information sharing. In other cases, such as the identity or other personal information relating to the staff involved in system development there may be legal barriers, including GDPR, to prevent information sharing, even if the SI wished to share this with the DP to bolster the credentials of their system.

\begin{figure}
\begin{center}
\includegraphics[width=0.35\textwidth]{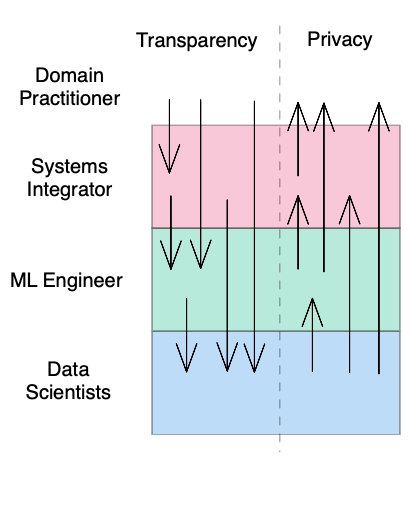} 
\end{center}
\vspace{-10pt}
\caption{Transparency and Privacy Tensions in an AI System}
\vspace{-5pt}
\label{needs}
\end{figure}

This tension between a desire for transparency versus a need for privacy in terms of information flows between parties with different motivations has been described by Trask, et al,~\cite{trask2020beyond} as \emph{structured transparency} and illustrated in Figure \ref{needs}.

\subsection{The ML Engineer}
The MLE is the role responsible for developing, testing and publishing the ML model. The MLE selects data to use to train the model and to test the model. This data can come from a number of sources, both within the MLE's own team or organisation, or can be sourced as secondary data from peers, commercially, or from open public sources, including Government or city authorities or global actors including UN or The World Bank~\cite{ismail2021ai}.

\subsubsection{Responsibilities}
The MLE has a responsibility to ensure that data used in the preparation of their ML models are fit for purpose, and will not cause unanticipated problems for users further up the hierarchy as they adopt and use the ML model in the development of AI systems or in the use of such systems. This places an onus on the MLE to take great care with the selection of data used. Beyond selecting data, the MLE has a responsibility to use the data ``correctly'' - that is, to use it in a manner that is consistent with the purposes for which the data was designed and shared, and in accordance with any terms and conditions set down in data licenses.

\subsubsection{Appropriate Data Requirement}
The degree to which an MLE can determine the suitability of data for use in the ML models they create and publish depends on their perception of the qualities of the underlying data or the party providing it. Information of this nature about the contributing data is assessed by the MLE across Trust Frontier C in Figure~\ref{fig:roles}.

\subsubsection{Privacy Requirement}
Repeating observations from the discussion of the SI's requirements, there is a tension between calls for visibility or transparency into the contributing elements and processes of ML models, and a need for protection of information about private individuals and commercially sensitive information belonging to the MLE and her partners and stakeholders.

\subsection{The Data Scientist}
The Data Scientist (DS) role is considered to involve the collection and generation of data for re-use by third parties\footnote{In some cases, the DS may offer data through an intermediary data broker \cite{immonen2014requirements}}. Such data is also known as secondary data. There is a relationship between the MLE and the DS, in that the DS needs to prepare data such that the MLE is willing and able to use the data provided.

There is much historical and ongoing debate in literature and policy about how to motivate a DS to make data available for use by third parties, with a particular concern being that there is often no direct incentive for them to do so\cite{norris2019science}. That line of research is largely outside of the scope of this paper, which focuses primarily on the needs of the MLE in adopting shared, secondary data., however there are choices made by the DS that have impacts that travel up the hierarchy, and can impact the ML model and AI system as ultimately perceived and used by DPs. Note that \emph{Data Governance}~\cite{khatri2010designing} is an area that has received attention, and is where frameworks for addressing these responsibilities and requirements would lie.

\subsubsection{Responsibilities}
The DS has a responsibility to ensure that data is fit for purpose, and will not cause problems for users further up the hierarchy as they make use of the products derived from the data. This places an onus on the DS to take care in data preparation, and in accurately documenting their data. Sambasivan, et al.~\cite{sambasivan2021everyone} provide many examples of problems caused by ``Data Cascades'' in AI systems, as the effects of brittle or poor quality data ripples through to the domain. This is not to say that responsibility for avoiding issues lies solely with the originating DS -- there is an onus on other roles to make sure that data is used sensitively, and within the bounds of which it is intended. The DS can facilitate this by preparing and documenting the data accurately and purposefully.

\subsubsection{Privacy Requirement}
There is a tension between calls for transparency about data used in AI systems and a desire for privacy. The privacy of data can be considered in two ways, first as we have seen above, there may be a research or business requirement from the DS to keep details of datasets and the processes of data generation and curation secret, in order to protect their intellectual property. Further, there may be a requirement, from GDPR for example, to protect the personal information about the staff involved in the production of data assets --- the qualifications of a DS, for example. There are also significant potential implications surrounding the privacy and traceability of individual data items within the dataset, especially when the data is about people.

\section{Requirements Modelling}
\label{chapter:modelling}
\begin{figure}[ht]
\centering
\includegraphics[width=0.45\textwidth]{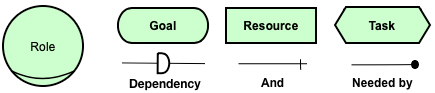}
\caption{Key iStar 2.0 Symbols, from Yu\cite{yu2011modeling}} \label{fig0}
\end{figure}

Graphically modelling sociotechnological systems provides a helpful way to visualise the roles and actors in complex systems, and identifying the relationships between participants. Literature from the field of Actor Based Modelling offers a useful starting point, and the iStar framework developed by Yu\cite{yu2011modeling}, and subsequently by Dalpiaz et al.\cite{dalpiaz2016istar}, provides metaphors to define actors and roles in a system, and to specify the goals of each role, as well as tasks they need to perform, and resources they have access to (summarised in Figure \ref{fig0}). iStar also allows dependencies to be specified, showing goals that actors need other actors to attain, tasks they are dependent on other actors to perform or resources they need other actors to provide them with.

\begin{figure*}[ht]
\centering
\includegraphics[width=1.0\textwidth]{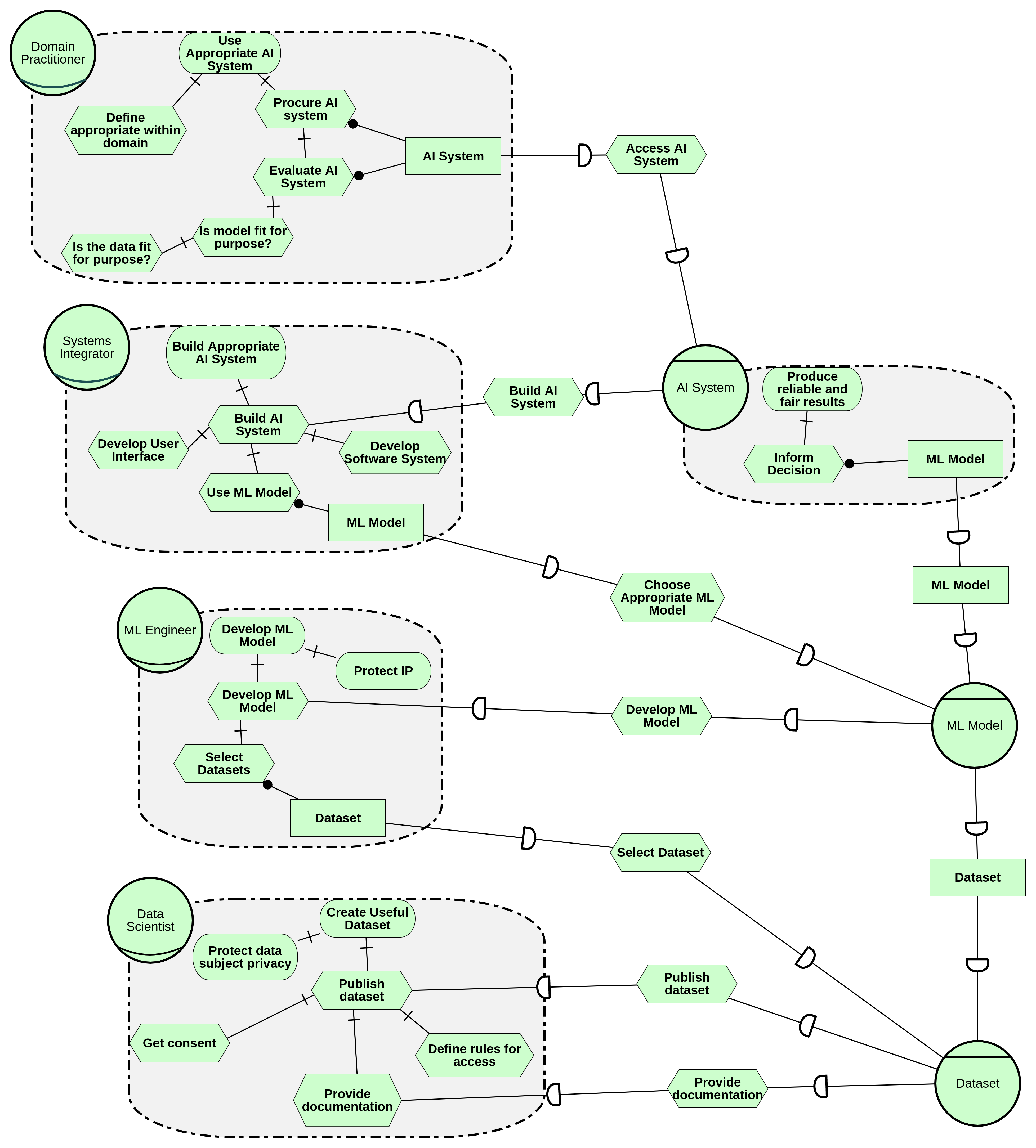}
\caption{Roles and Dependencies in AI Systems} \label{istarview}
\end{figure*}

Figure \ref{istarview} shows an iStar diagram for an AI system from the perspective of a DP, with a stated goal of ``Use an Appropriate AI System''. The model shows task dependencies which need to be satisfied, which includes defining the meaning of \emph{appropriate} in their domain, as well as evaluating a system, which has dependencies on many other roles and entities in the ecosystem.

\section{Discussion}
\label{chapter:discussion}
The pipeline outlined in this work defines a generic set of roles, responsibilities and requirements for actors involved in the development and application of an AI system within a domain. Each domain will come with it's own unique requirements for actors fulfilling these roles. The ultimate responsibility, and hence risk, for deploying these systems rests with the DP who will use it. In order to take on this risk, the DP needs assurance that the system is applicable, accurate and reliable for the use case it is applied. They need to trust it, and be able to demonstrate to others that it is trustworthy. This is especially important when the AI system is influencing decisions that impact human actors. For example in healthcare or policing systems.

Before trust can intelligently be placed in an AI system, the DP requires information about the systems constituent parts. Information sought might include details on how components were developed, and how, and how they were combined and integrated and by whom? However, complete transparency is unlikely to be desirable or achievable due to privacy and confidentiality concerns. Rather the DP needs to be able to verify specific claims about the system they are using, and ideally claims provided by sources who were involved in it's construction and whose trustworthiness can also be assessed.


In framing privacy, Nissenbaum determines that the flow of information between actors in different roles is regulated by contextual integrity ---  social contexts that influence the appropriate levels of information flow between the parties in their roles, defining \textit{context-relative informational norms}~\cite{nissenbaum2020privacy}. Identifying the roles is an important aspect of specifying the informational norms for any context, and determining what are considered to be appropriate expectations for information flows between them. To some degree, this describes the tension uncovered in this work between the DP need for visibility onto components and their constitution and the desire --- or necessity, in some instances --- for privacy from roles down the hierarchy, the SI, MLE and DS. Who each have different motivations to both share and withhold information, depending on context. This has been considered in some regards within an AI context by Trask, et al,~\cite{trask2020beyond} as \emph{structured transparency}, although this work is concerned more with the details of contents of datasets, rather than the less granular, hierarchical view of AI systems developed here. Beyond roles, the context of an AI system and establishment of information sharing norms between the actors in its ecosystem is heavily dependant on the circumstances in which the instance of the system is deployed --- with far greater need for integrity among actors being required in a healthcare deployment than a playlist algorithm, for example. 

\section{Conclusions}
\label{chapter:conclusions}
As AI systems are increasingly being implemented and integrated into the fabric of society, developing better practices around their development and deployment is critical. Despite a plethora of evidence pointing to biases encoded into some algorithms by their designers and the data used to train them \cite{o2016weapons,benjamin2019race,buolamwini2018gender}, they are being used in hiring processes \cite{brown2011writing}, credit scoring \cite{wei2015credit} and policing \cite{bennett2018algorithmic}. Protecting citizens from such systems is a shared responsibility, but on a day-to-day level, citizens have no choice but to place their confidence in the hands of DPs to have adopted suitable tools for their work. DPs have an unenviable task, but without reliable information on the constitution of AI systems, it is made almost impossible.

Clarifying the different roles and the hierarchy of interlinked dependencies between these roles in an AI system helps to define expectations and establish a specification for an information sharing environment among actors, where often there is no direct relationship. This highlights the tensions between the desires for transparency from SIs, MLEs and DPs, and the requirement for varied levels of privacy from SIs, MLEs and DSs themselves on the assets they have produced and contributed. An environment that seeks to meet the diverse and conflicting motivations and requirements of each role will need to offer a means to support end-to-end provision of trustworthy information. This will need flexibility to support granular, and deployment specific, flow of trustworthy information exchange between ecosystem participants. Central to this requirement is a need for the DP to be provided with support and tools to enable them to make assured decisions on the appropriate application, or abandonment, of a particular AI system for their domain, based on trustworthy and useful information being provided to them from all contributors to aid in their vital decision making.

\clearpage
\clearpage

\bibliographystyle{ACM-Reference-Format}
\bibliography{sample-base}
\end{document}